\documentstyle[prd,aps,psfig]{revtex}
\def\geff{\gamma_{\rm eff}}
\def\mpl{m_{\rm pl}}
\def\beq{\begin{equation}}
\def\eeq{\end{equation}}

\begin{document}

\draft

\renewcommand{\topfraction}{0.99}
\renewcommand{\bottomfraction}{0.99}
\renewcommand{\thefootnote}{\alph{footnote}}


\title{Extended Warm Inflation}

\author{J. M. F. Maia$^{1,2}$\footnote{jmaia@dfte.ufrn.br} and J. A. S. Lima$^{1}$ \footnote{limajas@dfte.ufrn.br}}
\address{1. Universidade Federal do Rio Grande do Norte \\
Departamento de F\'\i sica Te\'orica e Experimental, C. P. 1641, \\59072-970, Natal, RN, Brazil.}
\address{2. Universidade de S\~ao Paulo, Instituto de F\'\i sica, FINPE, CP 66318\\
 05389-970, S\~ao Paulo, SP Brazil. }

\maketitle

\begin{abstract}

A bidimensional parameter space of unitary area is introduced to study
phenomenologically the dynamic and thermodynamic behavior of inflationary models driven by a scalar field coupled to a thermal component. In this enlarged context the isentropic and warm pictures are just extreme cases of an infinite two-parametric family of possible inflationary scenarios. It is also shown that strong couplings are responsible by
an alteration in the standard slow roll conditions which relax the smallness constraint on the first and the second derivatives of the potential. 


\end{abstract}

\section{Introduction}

	Since the advent of inflationary models, the roles played by scalar
fields at different epochs of the cosmic history have been extensively
investigated. Such fields have been invoked in a variety of disparate
scenarios with rather different goals.  Some known examples are: (i) the
inflaton, the field that drives inflation \cite{Guth}, (ii) the axion, a cold
dark matter candidate \cite{KolbTurner} and (iii) the dilaton, the field
appearing in the low energy string action \cite{Green} which addresses the
same issues of inflation and may provide a solution to the singularity
problem \cite{GasperiniRRJ}. More recently, inspired by the existing
observational data and theoretical speculations, some authors have also
suggested scalar fields (sometimes called ``quintessence'') as the sough
non-baryonic dark matter\cite{Ratra}. These ``remnant" fields, might have
important consequences on the formation of the large scale structure, as well
as be responsible by the present day accelerated phase \cite{quintessence},
as indicated by the latest type Ia Supernovae observations \cite{Perlmutter}.

	Despite their generalized use in the cosmological framework, the
physical situations in which the scalar fields are commonly considered are
rather particular. For example, in the new inflation case there is not a
fundamental justification for ``turning off'' all the possible couplings
during the slow rollover phase and doing the opposite just at the onset of
the thermalisation phase of reheating (as well as afterwards, if some
potential energy is still available). This can be achieved only by
considering very strict initial conditions, which weakened the viability of
such a scenario. But this is not an isolated case.  Neglecting possible
thermal couplings between scalar fields and the other constituents of the
universe is a general feature assumed in nearly all scalar field models
presented in the literature.

	One important exception is the explosive reheating period required to succeed any version of isentropic inflation. Along this process, as the field coherently oscillates about the minimum of the potential, its energy is drained to the matter and radiation components.  Either in its earlier
\cite{Oldreheat} or in its modern version based on parametric resonance
(sometimes called preheating)\cite{Newreheat,Kofman}, the reheating
mechanism is a relatively fast process, and virtually all the entropy of the
present universe may be generated in this way. This is certainly not
the most general case. In principle, a permanent or temporary coupling of the scalar field $\phi$ with other fields might also lead to dissipative processes producing entropy at different eras of the cosmic evolution.  It is expected that progresses in non-equilibrium statistics of quantum fields will
provide the necessary theoretical framework for discussing dissipation in
more general cases (see for example \cite{Boyanovskyetal} and references
therein). Another possibility is the so-called ``instant preheating''
\cite{FKL}. In this process, the inflaton decays continuously into another
scalar field as it rolls down the potential. This second field is very
short-lived and rapidly decays into fermions thus furnishing a sustained
entropy generation, including for quintessence-like models.

	Although a justification from first principles for dissipative effects
has not been firmly achieved, such effects should not be ruled out only on
readiness basis. Much work can be done in phenomenological grounds as, for
instance, by applying nonequilibrium thermodynamic techniques to the problem
or even studying particular models with dissipation. An interesting example
of the latter case is the warm inflationary picture recently proposed \cite{Berera}. Like in new inflation, a phase transition driving the
universe to an inflationary period dominated by the scalar field potential is
assumed.  However, a standard phenomenological friction-like term
$\Gamma\dot\phi^{2}$ is inserted into the scalar field equation of motion to
represent a continuous energy transfer from $\phi$ to the radiation
field. This persistent thermal contact during inflation is so finely 
adjusted that the scalar field evolves all the time in a damped regime
generating an isothermal expansion. As a consequence, the subsequent reheating
mechanism is not needed and thermal fluctuations produce the primordial
spectrum of density perturbations \cite{Bererafang,Leefang}(see also reference \cite{Bellini}).

	Warm inflation was originally formulated in a phenomenological setting,
but some attempts of a fundamental justification has also been presented
\cite{BGR}. Furthermore, a dynamical systems analysis \cite{HR} showed that a smooth transition from inflationary to a radiation phase is attained for many values of the friction parameter, thereby showing that the warm scenario may be a workable variant to inflation. As it appears, its unique negative aspect is
closely related to a possible thermodynamic fine-tunning, because an
isothermal evolution of the radiative component is assumed from the very
beginning in some versions of warm inflation (for comments on this issue, see
\cite{BGR}). In other words, the thermal coupling acting during inflation is
so powerful and finely adjusted that the scalar field decays ensuring a
constant temperature even considering the exponential expansion of the
universe.

In brief, the aim of this paper is to relax this hyphothesis.  However,
instead of proposing another particular inflationary model, we discuss how
the differences between the isentropic and the isothermal inflationary scenarios can be depicted in a convenient parameter space. As we shall see, these models are only two extreme cases of an infinite two-parametric family. Hopefully, this unified
view may indicate ways to a consistent phenomenological treatments of these
models based on the methods of nonequilibrium thermodynamics.  We also
discuss how the standard slow roll conditions are modified due to the scalar
field decay.

\section{Scalar Field with Dissipation}

	We will limit our analysis to homogeneous and isotropic universes,
described by the flat Friedmann-Robertson-Walker (FRW) line element
\beq
ds^2 = dt^2-a^2(t)\left(dr^2  + r^2d\theta^2 + r^2\sin^2\theta d\phi^2\right), 
\eeq
where $a(t)$ is the scale factor (in our units $\hbar = c = 1$). The source
of this spacetime is a mixture of a real and minimally coupled scalar field interchanging energy
with a perfect fluid representing all the other fields. The Lagragian density
for the scalar field is
\beq
{\cal L} = {1\over 2}\partial^{\mu}\phi\partial_{\mu}\phi - 
		V(\phi) + {\cal L_{\rm int}},
\eeq
where the interaction is implied by the term $\cal L_{\rm int}$ and $V(\phi)$
is the scalar field potential. This field has the stress-energy tensor given
by
\beq
T^{\mu\nu}_{\phi} = \partial^{\mu}\phi\partial^{\nu}\phi - {\cal L}g^{\mu\nu}.
\eeq

The other component of the mixture is a simple fluid with energy-stress
tensor
\beq
T^{\mu\nu}_m = (\rho + p)u^{\mu}u^{\nu} - pg^{\mu\nu},
\eeq
where energy density and pressure are given respectively by $\rho$ and $p$.
The total energy stress tensor of the system $T_t^{\mu\nu} = T^{\mu\nu}_{\phi}
+T^{\mu\nu}_m$ obeys Einstein's field equations, from which we obtain the
equations of motion
\beq \label{friedmann00}
3H^2 = {8\pi \over \mpl^2}\left({\dot\phi^2 \over 2} + V(\phi) + \rho\right),
\eeq 
\beq\label{friedmannii}
3H^2 + 2\dot H = - {8\pi \over mpl^2}\left(
				{\dot\phi^2 \over 2} - V(\phi) + p\right),
\eeq
where a dot means time derivative, $H = \dot a / a$ is the Hubble parameter, $\mpl^2 = 1/G$ is the Planck mass, and we have used that the scalar field energy density and pressure are, respectively 
\begin{eqnarray}
&& \rho_{\phi}={1\over 2}\dot\phi^2 + V(\phi)\\
&& p_{\phi}= {1\over 2}\dot\phi^2 - V(\phi).
\end{eqnarray}
Now, assuming that the perfect fluid complies with the ``gamma law'' equation
of state,
\beq
p=(\gamma -1)\rho,
\eeq
the energy conservation law for this interacting selfgravitating mixture can be cast in the form
\beq\label{conserva}
\dot\phi (\ddot\phi + 3H\dot\phi + V'(\phi )) = -\dot\rho -3\gamma H \rho .
\eeq
where the prime denotes derivative with respect to the field $\phi$.
In order to decouple the two sides of the above equation and allow for the
decay of the scalar field into ordinary matter and radiation, we will assume
the usual phenomenological friction-like term (3$\Gamma\dot\phi^{2}$), thereby spliting (\ref{conserva}) in two  equations
\beq\label{phiGamma}
\ddot\phi + 3H\dot\phi + V'(\phi )= -3\Gamma \dot\phi
\eeq
and
\beq\label{rhoGamma}
\dot\rho + 3\gamma H \rho = 3\Gamma \dot\phi^2,
\eeq
where the dissipative coefficient $\Gamma$ is the decay width
of the scalar field and the factor 3 has been introduced for mathematical
convenience. For models endowed with an isentropic inflationary stage, we know that $\rho \sim 0$, at the onset of the
coherent oscillations, that is, when the field $\phi$ oscillates about the minimum of its potential $V(\phi)$. In these
cases, the term $\Gamma \dot\phi^2$ is generally inefficient for describing the first stages of the reheating process \cite{Newreheat,Kofman} and is not acting during exponential inflation. However, in the presence of a nonnegligible thermal
component, a friction term can be justified under special conditions during a de Sitter regime \cite{Boyanovskyetal}. Aplications of this theory lead to
complicated but promising warm inflationary models, with an unreasonably large quantity of light fields coupled
to the inflaton \cite{BGR}, although some of these models might be physically interpreted in terms of string theory\cite{BK99}. 

In what follows, it will be assumed the validity of this friction term as representing the thermal contact between $\phi$ and the other fields for any epoch. We also consider the ``thermal decay width" $\Gamma$ as a generic function of the temperature, or equivalently, of the cosmic
time.

Let us now define the following dimensionless parameters:
\beq\label{x}
x\equiv {\dot\phi^2 \over \dot\phi^2 + \gamma\rho}, 
\eeq
and
\beq\label{alpha}
\alpha \equiv {3H\dot\phi^{2} \over 3H\dot\phi^{2} + |\dot\rho|}.
\eeq
The adoption of these parameters is somewhat natural and may be understood as
follows. The warm picture departs quantitatively from
new inflation because during the inflationary stage the energy density of
the material component is not negligible. More precisely, it is not
negligible only in comparison to the scalar field kinetic term since the
potential $V(\phi)$ must be strongly dominant in order to generate
exponential expansion.  This means that one needs to compare the kinetic term 
$\dot\phi^{2}=\rho_\phi + p_\phi$  with a quantity involving the energy of
the material component written in a convenient form. For the mixture, the
(total) equivalent quantity is $\rho_t + p_t = \dot\phi^2 + \gamma\rho$. In
this way, recalling the energy conservation law, $x$ is defined by the ratio
between two ``inertial mass'' terms which redshift away due to the universe expansion. Additionally, it is not enough to compare the values of $\dot\phi^2$ and
$\gamma\rho$ since we also need to quantify how they evolve in the course of
the expansion. This explains the introduction of the parameter $\alpha$
involving  $3H\dot\phi^2$ and $\dot\rho$.  The presence of $|\dot\rho|$ in
this parameter is also reasonable. It comes into play because we are
adopting the original isothermal scenario proposed by Berera as a limiting case. Indeed, it seems to be an extreme theoretical situation where the cooling rate of the radiation due to expansion is fully compensated by the
transference of particles from the scalar field to the ordinary
constituents. It is implicit in the definition of $\alpha$ that $\dot\rho
\leq 0$, since a reheating phase is unecessary in warm inflation.
However we should point out that a positive $\dot\rho$ is possible and has
been investigated \cite{Leefang}.

The convenience of these parameters is apparent since by construction
they are dimensionless and constrained on the intervals $0 \leq x \leq 1$ and
$0 \leq \alpha \leq 1$. In particular, for the isothermal inflation
\cite{Berera,Bererafang} we have $\alpha=1$ and $x\rightarrow 0$, because $\dot\rho=0$
and $\gamma\rho \gg \dot\phi^{2}$.  For isentropic inflation one has $x=1$ and
$\alpha=1$ since $\gamma\rho \ll \dot\phi^{2}$ and $\dot\rho \ll
3H\dot\phi^{2}$ (the radiation becomes exponentially negligible). Similarly, noninteracting quintessence-like models lie at some intermediary value of
$\alpha = x$ between 0 and 1. When both parameters are equal to zero, we have the standard model (with a possible cosmological constant). Therefore, the most common solutions, with or without thermal couplings, can be portrayed in this bidimensional parameter space (see Fig. 1).

\begin{figure} 
\vspace{.2in} 
\centerline{\psfig{figure=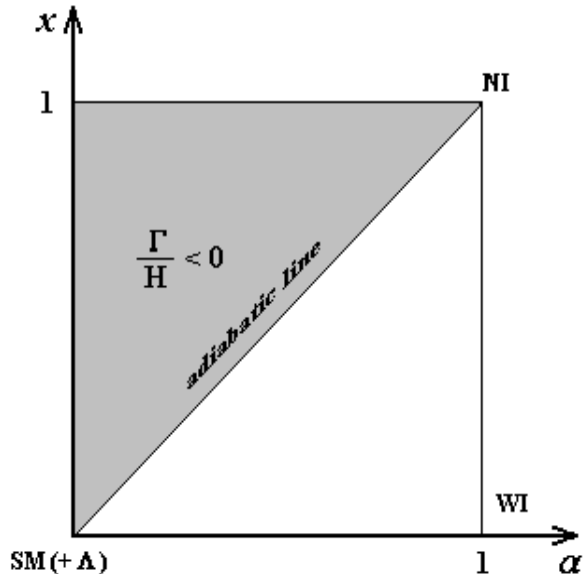,width=3truein,height=3truein} 
\hskip 0.1in} 
\caption{Bidimensional parameter space ($\alpha, x$) depicting thermal
couplings for the most common scalar field models with an arbitrary
potential $V(\phi)$. The shadowed overside triangle is forbidden for an
expanding universe. The entire diagonal line corresponds to the absence of
thermal couplings between $\phi$ and the other fields. The special points 
(0,0) and (1,1), represent the standard FRW model plus a cosmological constant
and the new inflation, respectively. The point WI (0,1) is the isothermal warm
inflationary scenario as originally proposed by Berera [13]. The
surprising feature is the large area available in this parameter space where
each point represent a possible inflationary evolution.} 
\end{figure}

\section{Inflation, Dissipation and Slow Roll Conditions}

A fundamental ingredient of many inflationary variants is the period of
``slow roll" evolution of the scalar field during which the inflaton field
evolves so slowly that its kinetic term remains always much smaller than
$V(\phi)$. A full description of the two-component mixture requires the
introduction of a dynamical parameter which is function of the total energy
density of the mixture. In order to discuss slow roll dynamical conditions, a
suitable choice of such a paramater is $\geff$, derived from manipulations on
Eqs.  (\ref{friedmann00}) and (\ref{friedmannii})
\beq\label{geff}
\geff = -{2\dot H \over 3H^2} = {\dot\phi^2 + \gamma\rho \over \rho_t }.
\eeq	
Slow roll conditions are imposed to assure (nearly) de Sitter solutions
for an amount of time, say, $\Delta t_I$, which must be long enough to solve
the problems of the hot big bang. By inspection of Eqs.  (\ref{friedmann00}),
(\ref{friedmannii}) and (\ref{geff}) one obtains that the condition for
having a de Sitter universe is
\beq\label{inequal}
3H^2 \gg 2|\dot H| \Leftrightarrow \geff \approx 0
		\Leftrightarrow V(\phi ) \gg \rho + {\dot\phi^{2} \over 2}.
\eeq
Additionally, a sufficient amount of inflation requires the above relations
to be valid along the interval $\Delta t_I$. This usually implies constraints
on the slope and curvature of $V(\phi )$. However, as we shall see, the
coupling between $\phi$ and the $\gamma$-fluid during inflation may relax
these constraints for a large range of intermediary situations between the
isentropic and warm scenarios. In our case, these conditions are represented by (see
also\cite{Berera})
\beq\label{src1}
H^2\approx {8\pi\over 3\mpl^2}V(\phi),
\eeq

\beq\label{src2}
3(H + \Gamma)\dot\phi \approx - V'(\phi).
\eeq
Notice that the discussion that follows is independent of constraints on the
signal of $\dot\rho$, so that we are not limited to the definition of
$\alpha$. The approximated expressions (\ref{src1}) and (\ref{src2}) give us
the following constraints on the shape of $V(\phi)$ \beq
\left({V'(\phi ) \over V(\phi )}\right)^2{\mpl^2 \over 16\pi} \approx 
	\left(1+ {\Gamma \over H}\right)^2 {\dot\phi^2 \over 2V(\phi )}
\eeq
and
\beq
{V''(\phi ) \over V(\phi )} {\mpl^2 \over 24\pi} \approx 
	- \left(1 + {\Gamma \over H} \right) 
	{\dot\phi^2 + \gamma\rho \over 2V(\phi )} - {1 \over 3H}{d \over dt} 
	\left({\Gamma \over H}\right),
\eeq
We recall that the standard slow roll conditions appearing in isentropic
inflation are recovered for $\Gamma = 0$ (decoupled mixture). They imply
that inflation is possible only if the potential is extremely flat, by the
last inequalty in (\ref{inequal}).  However, in the limit ($\Gamma \gg H$),
one may see that the first and the second derivative of the potential are not
necessarily small, in order to guarantee the continued accelerated expansion.
Thus, the extremely flat potential of usual inflation assuring the
slow down of the scalar field by an enough amount of time, may be replaced by
a large friction-like term with no extreme behavior of the  derivatives of
$V(\phi)$.  The field does not accelerate, thereby $\dot\phi^{2}$ becoming
comparable to $V(\phi)$, because the friction may be really large and not
because the potential is unusually flat. This explains why this class of
extended scenarios may provide a solution for the fine tunning problems plaguing the new
inflationary picture.

\section{A Toy Model}

The standard procedure in scalar field cosmological model building
is to assume a specific potential $V(\phi)$, motivated or not from particle
physics. This potential is used to solve the $\phi$ equation of motion (\ref{phiGamma}). Subsequently, if there is a nonegligible energy density stored in the other fields, the solution of $\phi$ is inserted into (\ref{rhoGamma}), which is solved for the energy density  $\rho$. As widely known, even for a given coupling, many solutions are possible just changing the potential $V(\phi)$. To fix terminology, this method for generating solutions will be called dynamic approach. 

On the other hand, the parameter space ($\alpha, x$) can also be used as a guide in the search for new models. To show that a different (thermodynamic) route is also possible we first rewrite (\ref{rhoGamma}) in terms of $x$
and $\alpha$. We have
\beq\label{Gamma/H}
{\Gamma \over H} = {1 \over x} - {1 \over \alpha}.
\eeq
It is worth noticing that (\ref{Gamma/H}) does not include the potential
since it is representative only of the possible couplings, but not of the
dynamics. In what follows we obtain a
simple model example based on equations (\ref{x}), (\ref{alpha}) and (\ref{Gamma/H}). In principle, a more quantitative analysis will clarify the interesting features contained in the parameter space ($\alpha, x$). In order to be as generic as possible, we also include the possibility of a decaying $\phi$ during any era.

Since $x$ and $\alpha$ do not depend on $V(\phi)$, one may show that equation (\ref{rhoGamma}) for the material medium can be rewritten as  
\beq\label{dotrho}
\dot\rho = -3\gamma  H \rho \left(1-\delta\right),
\eeq
where for short we have introduced the quantity 
\beq\label{delta}
\delta\equiv {\alpha - x \over \alpha (1-x)} .
\eeq 
It is apparent that the simplest toy model is the one where the parameters $\alpha$ and $x$ are constants. In this case, the solution of (\ref{dotrho}) is
\beq\label{rhosol}
\rho (a) = \rho_{\rm I} \left({a \over a_{\rm I}}\right)^{-3\gamma
		(1 - \delta )} ,
\eeq
where the integration constants, $\rho_{\rm I}$ and $a_{\rm I}$, are  ``initial" conditions just at the onset of the inflationary stage. It should be noticed that if $\delta=1$, or equivalently, if $\alpha=1$, the energy density $\rho$ remains constant ($\rho=\rho_{\rm I}$). In particular, if $x \rightarrow 0$, that is, if $\rho \gg {\dot{\phi}}^2$, we recover the isothermal inflationary scenario proposed by Berera (see also Fig.1). This show more clearly why models with values of $0 < x < 1$ may also evolve isothermally during inflation (exponential or power-law). The other extreme situation is obtained if $\delta =0$, that is, if $\alpha=x$, with the energy density scaling as $\rho \sim a^{-3\gamma}$. This behavior is typical for an adiabatic expansion (very weak coupling), and in Fig. 1, it corresponds to the adiabatic line. Note that even in these circumstances, where the parameters are constants, one may expect that an intermediary situation ($0< \delta < 1$) be physically more probable. 

Under the above conditions, the energy density of the the scalar field may also be readily obtained. First we rewrite (\ref{phiGamma}) as 
\beq\label{rhophiGamma}
\dot\rho_{\phi} = -3H\dot\phi^2 \left(1 + {\Gamma \over H}\right) .
\eeq
Now, since $x={\rm const}$ we see from (\ref{x}) that
$\dot\phi^2 = {x\over 1-x}\gamma\rho$, and using (\ref{Gamma/H}) and (\ref{rhosol}), this equation can be
integrated in terms of the scale factor. The result is
\beq\label{rhophisol}
\rho_{\phi} = \rho_{\phi_I} + \rho_I\left(\delta + {x \over {1 - x}}\right)\left[{\left({a \over a_I}\right)^{{-3\gamma}(1-\delta)} - 1} \over {1 - \delta}\right] ,
\eeq
where the constant $\rho_{\phi_I}$ is the scalar field energy density when $a=a_I$. 

In the adiabatic limit ($\delta=0$) the above expression reduces to
\beq\label{rhophisol0}
\rho_{\phi} = \rho_{\phi_I} + \rho_I{x \over {1 - x}}\left[{\left({a \over a_I}\right)^{-3\gamma} - 1}\right] ,
\eeq
The isothermal case ($\delta = 1$) is also readily obtained using that 
$\lim_{q \rightarrow 1}{f^{1-q}-1\over 1-q} = \ln f$. One finds 
\beq\label{delta1}
\rho_{\phi} = \rho_{\phi_I} -\frac{3\gamma}{1 -x}\rho_I \ln\left({a \over a_{\rm I}}\right)
\eeq
If $\delta \neq 1$, using ($\ref{rhosol}$) and ($\ref{rhophisol}$), we have that the total energy density may be written as
\beq\label{rhototal}
\rho_{t} = \rho_{\phi_I} - B\rho_I + B\rho
\eeq
where $B={x(1 + {\Gamma \over H}) \over (1 - x)(1 - \delta)}$ is a constant.

	The time dependence of the scale factor may also be obtained from (\ref{x}) and (\ref{geff}), which give $\geff = {\gamma \rho \over (1 - x)\rho_t}$. We see that the possible evolution laws depend critically on the initial conditions. In particular, if $B \sim {\rho_{\phi_I} \over \rho_I}$ we have 
$\geff \sim {\gamma \rho_I \over (1 - x)\rho_{\phi_I}}$. Thus, if $\rho_{\phi_I} \gg \rho_I$ we have exponential inflation, and if $\rho_\phi$ dominates only moderately, the scale factor will evolve as a power law inflation with a coupled thermal component. As one may check, in this case  the potential $V(\phi)$ scales as $e^{-\lambda \phi}$, where $\lambda(\delta)$ is a positive parameter.  Note still that if the first two terms in (\ref{rhototal}) do not cancel each other, we may have a ``remnant" cosmological constant for large values of the cosmological time. 

As we have seen, this simple model is representative of two different ways of extending the original warm inflation. One is keeping the isothermal condition ($\alpha = \delta = 1$) for generic values of $x$. Another approach is to consider nonvanishing parameters $\alpha$ and  $x$ ($\alpha\neq x$), which allows for warm power law inflationary models.  Hopefully, in a more general framework, where the pair of parameters ($\alpha, x$) are time dependent functions, a consistent unified picture containing the new and warm inflation, as well as all the intermediary situations may be obtained.

\section{Conclusions}

Our analysis might be useful as an heuristic tool for building models
with scalar fields coupled to a material medium. In principle, it can be
applied to warm inflation-like models (with or without phase transitions
involved), reheating (even for the old scenario) and quintessence-like
models. As a rule, it could be tried as a first step, before attempting
different potentials for the scalar fields. As we have shown, this happens
because depending on the two parameters, the conditions for distinct dynamics of the scale factor (including inflationary regimes) are relatively independent of the shape of the potencial, since a large dissipative term may provide a natural slow rolling for the field. This may unconstrain the conditions on the flatness of the potential in such a way that even exotic models like the oscillating inflation of Damour and Mukhanov\cite{Damour} may provide the necessary number of e-foldings and enough post-inflationary
radiation temperature. The toy model presented in the last section somewhat suggest that any inflationary interpolating solution between the isentropic and isothermal limits can be represented in the bidimensional parameter space ($\alpha, x$). Particular examples will be discussed elsewhere \cite{LM99}.

	As it appears, a more comprehensive phenomenological treatment of this matter should necessarily include thermodynamical constraints thus requiring the methods and techniques from
nonequilibrium thermodynamics.  As shown recently\cite{termoscalar},
a phenomenological coupling term explicitly dependent on the created
particles (and not only on the scalar field) should be a natural outcome of
these methods when the decay products thermalize with the heat bath. Such an approach may have interesting consequences on the old reheating and warm inflation models.

{\bf Acknowledgements} The authors are grateful for the support of CAPES and CNPq (Brazilian 
research agencies). One of us (JASL) was also supported by the project PRONEX/FINEP 
(No. 41.96.0908.00).

\end{document}